\documentclass[aps,twocolumn,prb,showpacs,superscriptaddress]{revtex4}
\usepackage{amssymb}
\usepackage{epsfig}
\usepackage{epsf}
\usepackage{subfigure}
\usepackage{amsmath}

\def\be{\begin{equation}} \def\ee{\end{equation}}
\def\bea{\begin{eqnarray}} \def\eea{\end{eqnarray}}

\def\nn{\nonumber}

\begin{document}

\title{Spectral Weight Transfer in Multiorbital Mott Systems}

\author{Wei-Cheng Lee}
\email{leewc@illinois.edu}

\author{Philip W. Phillips}
\email{philip@vfemmes.physics.illinois.edu}
\affiliation{Department of Physics, University of Illinois, 1110 West Green Street, Urbana, Illinois 61801, USA}

\date{\today}

\begin{abstract}

We develop here a general formalism for multi-orbital Mott systems
 which can be used to understand dynamical and static spectral weight
 transfer.
We find that the spectral weight transferred from the high energy
scales is greatly increased as a result of the multi-orbital
structure. As a consequence certainly dynamically generated symmetries
obtain at lower values of doping than in the single-band Hubbard
model.  For
example, in the atomic limit for filling less than one electron per site, the particle-hole symmetric condition
in the lower band shifts from the one-band result of $x=1/3$ to $x=1/(2n_o+1)$, where $n_o$
is the number of orbitals with an unpaired spin. Transport properties
computed from effective low-energy theories which forbid double occupancy of bare
electrons, such as the multi-orbital t-J generalization, should all be
be sensitive to this particle-hole symmetric condition.  Away from the atomic limit, the
dynamical contributions increase the transferred spectral weight.
Consequently, any phenomena which are sensitive to an emergent
particle-hole symmetry should obtain at $x<(1/(2n_o+1)$.  
\end{abstract}

\maketitle

\section {Introduction}

The parent of the copper-oxide superconductors is a single-orbital Mott insulator in which one unpaired
d-electron is responsible for the essential physics. By contrast, the iron
pnictides are multi-orbital bad metallic systems in which several
d-electrons play an active role.  Given that both systems
superconduct, it is highly suggestive that combining the key
ingredients of the two might result in a higher $T_c$ material.
One of the ubiquitous features of the cuprates is dynamical spectral weight
transfer\cite{cooper1990,uchida1991,chen1991} arising fundamentally from the lack of rigidity of the
Hubbard bands in a Mott system. Such spectral weight rearrangments are
well documented in optical conductivity measurements\cite{cooper1990,uchida1991} and oxygen K-edge
x-ray absorption\cite{chen1991}. Spectral weight rearangements in
the cuprates are
well described\cite{meinders1993,eskes1994} by the single-band Hubbard model.  Classic Mott systems such
as VO$_2$ and NiO contain several d-orbitals that are singly occupied.
At present, although there are some studies on Mott transitions in multiorbital systems using dynamical mean-field theory (DMFT)
\cite{rozenberg1997,han1998,ono2003,koga2004,jakobi2009}, there is no formalism that describes the spectral weight
rearrangements in such systems under doping nor any experiments
that observe this effect. 
In particular, since DMFT neglects the momentum-dependence in the single particle self-energy, the essential physics of 
hopping-induced spectral weight transfer under the introduction of dopings can not be captured correctly.  Such momentum dependence is particularly important in the treament\cite{XAS2011} of mixed-valent systems in which inter-site coherences dominate the physics.
It is precisely the theoretical
description of spectral weight rearrangement in multi-orbital Mott
systems that we tackle here.  
We develop the strong coupling perturbative analysis for multi-orbital
systems and derive expressions for the spectral weight shifts in the lower band as a
function of doping. The Hund's couling, a new energy scale absent in the single band Hubbard model, is included explicitly in this study.
It has been pointed out that Hund's coupling play a crucial role in determining the correlation effects\cite{medici2011}.
As a result, it is also important to undersatnd how Hund's coupling affects the spectral properties which is also investigated in this paper.
The key result we obtain is that multi-orbital
physics enhances the total spectral weight transferred across the Mott
gap. It is our hope that this study will motivate experiments on doping
multi-orbital Mott insulators.  

Indeed, the low energy physics of strongly correlated systems remains one of
the most challenging problems.
The difficulty in strong coupling is uncloaking a set of excitations
which get rid of the interactions.  The presence of spectral weight
transfer demonstrates that such entities, should they exist, are not
straightforwardly related to those in the UV-complete theory. For
example, it has been known since 1967\cite{harris1967} in the context of the 1-band
Hubbard model that the total spectral intensity of the lower band exceeds the
total number of electron states in the band.
Hence, the additional degree of freedom in the lower band must be
orthogonal to an electron.  The culprit that accounts for the excess
spectral weight in the lower band is dynamical spectral weight
transfer (DSWT)\cite{harris1967,eskes1994}.  DSWT emerges from mixing
between doubly occupied sites in the upper band with singly occupied
ones in the lower band.  The fact that this effect can account for the
orthogonality for electron addition at low energies implies that it
provides a natural mechanism to explain the vanishing of the
quasiparticle weight, $Z$, seen in angle-resolved photoemission\cite{fournier2010}
(ARPES) in the underdoped cuprates.  Consequently, it is important to
catalogue precisely how this effect is manifest in multi-orbital Mott
systems. We demonstrate here that this effect is even
more pronounced because the multi-orbitals enhace the phase space for
dynamical mixing between the electronic states below and above the
Mott gap.

This paper is organized as follows.  In Section II, we consider a
general multi-orbital Hubbard model in the presence of
Hund's coupling and analyze the spectral weight transfer for various electron fillings 
in the atomic limit. 
Due to the existence of Hund's coupling and the paired hopping with
the energy scales $J$, we show that the distribution of the spectral weights is enriched.
Moreover, the non-zero $J$ also complicates the calculation of
DSWT when the hoppings are introduced.   This is analyzed in Section III.
The full details of the derivations are placed in the Appendix.
\section{Spectral weight transfer in atomic limit}
The Hamiltonian describing a multi-orbital Mott system consists of two parts,
\be
H=H_K + H_I,
\label{hamfull}
\ee
with $H_K$ the non-interacting tight-binding Hamiltonian
\be
H_K = - \sum_{\langle ij\rangle a b\sigma} t_{ab}c^\dagger_{i a\sigma}c_{j b\sigma},
\label{hamk}
\ee
and $c^\dagger_{i a \sigma}$ creates an electron with spin $\sigma$ in the orbtial $a$ on site $i$.
Due to the existence of the multi-orbital structure, electrons on the same site experience different types of interactions, 
including the intra- and inter-orbital Coulomb repulsive interactions, Hund's coupling, and the paired hopping between different orbitals.
The resulting on-site interaction for the multi-orbital system reads
\bea
H_I &=& \sum_{ia} U n_{ia\uparrow}n_{ia\downarrow} + \sum_{i,b>a}(U'-\frac{J}{2}) n_{ia} n_{ib} \nn\\
&&- \sum_{i,b>a}2J\vec{S}_{ia}\cdot \vec{S}_{ib} + J\big(p_{ia}p_{ib}^\dagger + h.c.\big),
\label{hamI}
\eea
where $n_{i a \sigma}=c^\dagger_{i a\sigma}c_{i a\sigma}$, $n_{i a}=n_{i a\uparrow} + n_{i a\downarrow}$, 
$\vec{S}_{i a} = \frac{1}{2}c^\dagger_{i a \mu}\vec{\sigma}_{\mu\nu}c_{i a \nu}$, $p_{i a}=c_{i a\uparrow}c_{i a\downarrow}$.
It should be noted that $U'=U-2J$ holds for systems with both orbital degeneracy and cubic symmetry. 
This relation will be used in this paper unless stated otherwise.

\subsection{Two-orbital Mott system}
Before we present the result for the 2-orbital system, let us recount the atomic-limit result for the single-orbital Mott insulator.  In this limit, the exact single-particle retarded Green function is
\be\label{specweight}
G_{i}^R(\omega)=\frac{1+x}{\omega-\mu+\frac{U}{2}}+\frac{1-x}{\omega-\mu-\frac{U}{2}},
\ee
The two poles, split by $U$, carry
spectral weight of $1+x$
and $1-x$ respectively.  Each hole subtracts a single state from the
high energy sector, thereby leading to the $1-x$ residue.  The empty
state reappears at low energies as part of the addition spectrum at
low energies.  The fact that the empty part of the spectrum grows as
$2x$ as opposed to $x$ in a band insulator reflects this static
transfer of spectral weight. 

\begin{figure}
\includegraphics{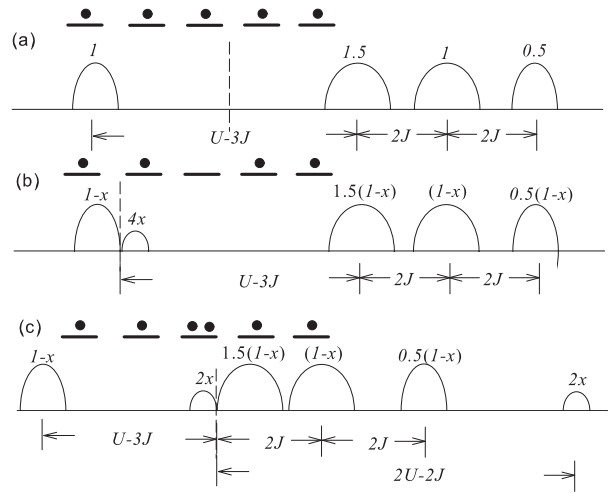}
\caption{\label{fig:one} The schematic illustration of spectral weight transfer in a two-orbital Mott system for (a) $n=1$, (b) $n=1-x$, and (c) $n=1+x$.
$U'$ has been replaced by $U-2J$ in the plot. The y-axis represents the magnitude of the spectral weight and the x-axis, the energy.  Although the bands are strictly $\delta-$function peaks in the atomic limit, they are represented as broadened here entirely for aesthetic purposes.  The dashed line represents the location of the chemical potential.}
\end{figure}

To begin our study of a multi-orbital system, we start with the two-orbital case since a variety of transition metal oxides with orthorhombic crystal structure have degenerate 
quasi-1D $d_{xz}$ and $d_{yz}$ bands which dominate in the low energy physics.
In this case, the Mott physics will be prominent near integer fillings of $n=1,2,3$. 
One interesting question is how is the spectral weight redistributed when such systems are doped with holes or electrons near these integer fillings.
In the atomic limit, the spectral weights can be obtained exactly by evaluating
\be
A^{\pm}_i(\omega) = \pm \frac{1}{\pi}\sum_{a\sigma} {\rm Im} G^{\pm}_{ia\sigma}(\omega),
\label{aw}
\ee
where
\be
G^{\pm}_{ia\sigma}(\omega) = \sum_m \frac{\vert\langle\psi^{N\pm 1}_m
\vert c^\dagger_{ia\sigma} (c_{ia\sigma})\vert\psi^N_G\rangle\vert^2}{\omega - E^{N\pm 1}_m + E^N_G \mp i\eta}
\ee
is the electron removal (addition) Green function on site $i$, and
$\{E^N_m\}$ and $\{\psi^N_m\}$ are the eigenenergies and eigenvectors
of $H_I$ with two orbitals
listed in Table \ref{table:one}.
Because a non-zero $J$ lifts the degeneracy of two-electron states on a single site,
the spectral weight also splits into several parts separated by the energy scales of $J$ whenever two-electron states are involved.
For example, if $J=0$ and $n=1$ the spectral intensity should be 3 for the 'upper Hubbard bands' and 1 for the 'lower Hubbard band' with a gap of $U$ between them. 
As $J$ is turned on, using Eq. \ref{aw} and Table \ref{table:one}, we have that
\bea
A^+_i(\omega) &=& \frac{3}{2}\delta(\omega-\mu-U'+J) + \frac{1}{2}\delta(\omega-\mu-U'-J) \nn\\
&+& \frac{1}{2}\delta(\omega-\mu-U+J) + \frac{1}{2}\delta(\omega-\mu-U-J),\nn\\
A^-_i(\omega) &=& \delta(\omega-\mu).
\label{apm-1e}
\eea
It can be seen that the upper Hubbard bands split into three parts with energies of $U-3J$, $U-J$, and $U+J$ ($U'=U-2J$ is used) with different spectral intensities as shown in
Fig. \ref{fig:one}(a).
As the system is doped with either holes or electrons,
the redistribution of the spectral weights occurs at all energy scales in a manner similar to that in single-band Hubbard model\cite{harris1967}.
Moreover, the amount of spectral weight transferred from the high energy down to low energy is even larger in multi-orbital Mott systems.
The reason is that the multi-orbital structure provides more ways to add or remove electrons.
In the following, we will analyze case-by-case the spectral weight distributions near different interger fillings.

\begin{table}[t]
\caption{A complete list of the eigenstates and eigenenergies of $H_I$ given in Eq. \ref{hamI} for two orbitals on site $i$.}
\centering
\begin{tabular}{c c c c}
\hline\hline
No. of electrons (N) & Eigenstate & Eigenenergy\\ [0.5ex]
\hline % inserts single horizontal line
1 & Any 1-electron state & 0  \\
\hline
2 & $\vert 1\uparrow,2\uparrow\rangle\,\,\,,\,\,\,\vert 1\downarrow,2\downarrow\rangle$ & $U'-J$  \\
  & $\frac{1}{\sqrt{2}}(\vert 1\uparrow,2\downarrow\rangle+\vert 1\downarrow,2\uparrow\rangle)$& $U'-J$ \\
    & $\frac{1}{\sqrt{2}}(\vert 1\uparrow,2\downarrow\rangle-\vert 1\downarrow,2\uparrow\rangle)$& $U'+J$ \\
      & $\frac{1}{\sqrt{2}}(\vert 1\uparrow,1\downarrow\rangle-\vert 2\uparrow,2\downarrow\rangle)$& $U-J$ \\
        & $\frac{1}{\sqrt{2}}(\vert 1\uparrow,1\downarrow\rangle+\vert 2\uparrow,2\downarrow\rangle)$& $U+J$ \\
	\hline
	3 & Any 3-electron state & $U + 2U' -J$\\
	\hline
	4 & $\vert 1\uparrow,1\downarrow,2\uparrow,2\downarrow\rangle$ & $2U + 4U'-2J$\\ [1ex]
	\hline
	\end{tabular}
	\label{table:one}
	\end{table}

The case of $n=1-x$ is the most prominent example to illustrate the effect of multi-orbital structure on spectral weight transfer.
For an ${\cal N}-$site system, there are ${\cal N}(1-x)$ sites with one electron and ${\cal N}x$ sites empty in the ground state.
For sites occupied by one electron, $A^\pm_i(\omega)$ are the same as
in Eq. \ref{apm-1e}.
For unoccupied sites, we find that
\bea
A^+_i(\omega) &=& 4\delta(\omega-\mu),\nn\\
A^-_i(\omega) &=& 0.
\eea
Therefore the intensity of the lower energy bands near the chemical potential is simply 
\bea
m^h(1-x)&\equiv&\frac{1}{{\cal N}}\sum_i \big[\int^\mu_{-\infty} d\omega A_i^-(\omega) + \int^{\Lambda}_\mu d\omega A_i^+(\omega)\big]\nn\\
&=&1-x+4x = 1+3x, 
\label{mlhb-1-x}
\eea
where $\Lambda$ is a cutoff energy scale demarcating the division between IR and UV scales.
One can realize Eq. \ref{mlhb-1-x} by counting the total number of
ways to remove and add electrons to the system without
resulting in an energy change.
While there is one way to remove an electron from an occupied site contributing $1-x$ to $m^h(1-x)$, there are four ways to add an electron to an empty site in a 
two-orbital system with spin $1/2$ leading to the remaining contribution of $4x$.
This is precisely the same argument used to understand the spectral
properties in the single-band Hubbard model leading to the well-known result (see Eq. (\ref{specweight}))
that the total weight of the lower band is $m_{LHB}=1-x + 2x=1+x$\cite{harris1967,meinders1993}, where
$x$ is the number of holes.
In fact, for $n=1-x$, the spectral weight can be obtained straightforwardly based on the same argument for $n_o$-orbital systems with 
spin $1/2$ in the atomic limit,
\be
m^h(1-x) = 1-x + 2n_ox = 1+(2n_o-1)x.
\ee
Clearly, the multi-orbital structure amplifies the spectral weight transfer from high energies down to the chemical potential even in the atomic limit.
\subsection{$n_o$-orbital Mott system}

The generalization of the spectral weight transfer for $n>1$ is complicated by Hund's coupling and paired hopping terms which disperse the spectral weights 
at energy scales on the order of $J$ in the lower and upper Hubbard bands.
Nevertheless, since the Mott insulating phase at integer filling requires a well-defined Mott gap $\Delta_{Mott}$ demarcating the lower and upper 
energy bands, it is expected that the spread of the spectral weight due to $J$ near the chemical potential occurs within the Mott gap.
In other words, we can always find proper cutoff energy scales $\mu\pm E^\pm$ such that $E^\pm\sim O(J) < \Delta_{Mott}$. 
Then for hole-doping near integer fillings in an $n_o$-orbital system, the intensity of the bands near the chemical potential can be evaluated using
\bea
m^h(n_i-x) &\equiv& \frac{1}{{\cal N}}\sum_i\big[\int_{\mu-E^-}^{\mu}d\omega A^-_i(\omega)\nn\\
&+& \int^{\mu+E^+}_{\mu}d\omega A^+_i(\omega)\big]\nn\\
&=&n_i(1-x)+ (2n_o-n_i+1)x\nn\\
&=& n_i + (2n_o-2n_i+1)x.
\label{mlhb-h}
\eea
The above result can be understood as follows. The ground state with
filling $n=n_i-x$ has a total fraction of $(1-x)$ sites occupied (per site) by $n_i$ electrons and 
a total fraction of $x$ sites occupied by $n_i-1$ electrons.
The processes with energies on the order of $J$ away from the chemical
potential correspond to removing electrons from sites
occupied by $n_i$ electrons and adding electrons into sites occupied by $n_i-1$ electrons.
Consequently, there are $n_i$ ways for electron removal and $(2n_o - n_i + 1)$ ways for electron addition.
Eq. \ref{mlhb-h} just accounts for the sum of these ways to add and remove electrons discussed above.

\begin{figure}
\includegraphics{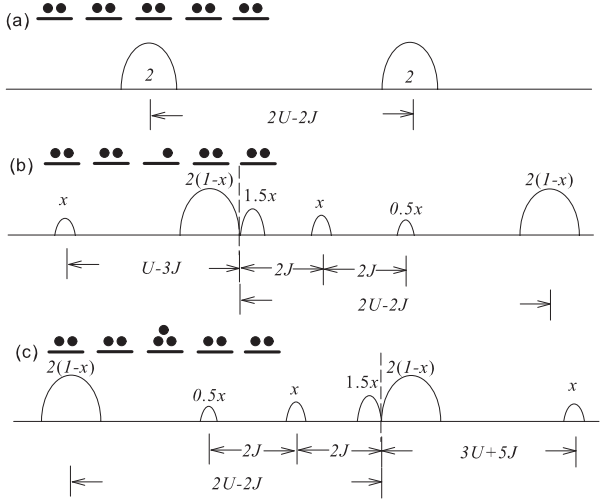}
\caption{\label{fig:two} Schematic illustration of spectral weight transfer in a two-orbital Mott system for (a) $n=2$, (b) $n=2-x$, and (c) $n=2+x$.
$U'$ has been replaced by $U-2J$ in the plot.  }
\end{figure}

Likewise, for electron-doping, $n=n_i+x$, the ground state is composed of ${\cal N}x$ of total sites occupied by $n_i+1$ electrons and
${\cal N}(1-x)$ of total sites occupied by $n_i$ electrons. 
As a result,  there are $n_i+1$ ways to remove electrons from sites
occupied by $n_i+1$ electrons and $(2n_o - n_i)$ ways to add electrons into sites occupied by $n_i$ electrons with 
energies of the order of $J$ away from the chemical potential. 
This yields
\bea
m^e(n_i+x) &=& \frac{1}{{\cal N}}\sum_i\big[\int_{\mu-E^-}^{\mu}d\omega A^-_i(\omega)\nonumber\\
 &+& \int^{\mu+E^+}_{\mu}d\omega A^+_i(\omega)\big]\nn\\
&=& (n_i+1)x + (2n_o - n_i)(1-x) \nn\\
&=& 2n_o - n_i + (1 + 2n_i -2n_o)x.\nn\\
\label{mlhb-e}
\eea
The spectral weights evaluated from Eq. \ref{aw} and Table \ref{table:one} for fillings of $n=1\pm x$ and $n=2\pm x$ in a two-orbital Mott system are illustrated 
in Figs. \ref{fig:one} and \ref{fig:two}. 
Proper choices of cutoff energies $(E^-,E^+)$ to account for the spectral intensity near the chemical potential $m^{e,h}(n=n_i\pm x)$ 
are $(-\infty,0^+)$ for $n=1-x$, $(0^-,4J)$ for $n=1+x$, $(0^-,4J)$ for $n=2-x$, and $(4J,0^+)$ for $n=2-x$ respectively.
It is interesting to mention that this generalized intensity of the bands near the chemical potential is symmetric with respect to 'half-filling' ($n_i = n_o$), i.e., 
\be
m^h(n_i-x) = m^e(2n_o - n_i + x).
\ee

\section{Dynamical pectral weight transfer}

In this section, we proceed to compute the dynamical spectral weight by treating $H_K$ as a small perturbation in the strong-coupling limit ($t_{ab}<<U,U'$).
The method used here is a generalization of the unitary
transformation method\cite{harris1967,macdonald1988,eskes1994,chernyshev2004} which has been successfully applied to study DSWT
in the single-band Hubbard model.
The formalism given below is general regardless of the magnitude of $U'=U-2J$.
Therefore, we will write $U'$ as an independent parameter from now on.
\subsection{General formalism of the perturbation theory}
The central idea of this method is to perform a unitary transformation
\be
H=e^{S}\tilde{H}e^{-S}\,\,\,,\,\,\,H=H_K + H_I\,\,\,,\,\,\,\tilde{H}=\tilde{H}_K + \tilde{H}_I
\label{ut}
\ee
with the condition
\be
[H,\tilde{H}_I]=0.
\label{constraint}
\ee
In other words, the goal is to find 'dressed' fermion operators $\tilde{c}_{ia\sigma}$ such that the original Hamiltonian $H$ shares the same 
eigen basis with the transformed interaction Hamiltonian $\tilde{H}_I$.
Since $\tilde{H}_I$ conserves the number of multiply occupied sites in terms of these dressed fermionic degrees of freedom, so does the original Hamiltonian $H$.
As a result, the ground state of $H$ can be expressed in terms of the number representation of the dressed fermionic degrees of freedom.
In general, for any operator $O$, we define $\tilde O$ such that
\be
O\equiv {\bf O}(c), \quad \tilde{O}\equiv {\bf O}(\tilde c),
\ee
simply by replacing the Fermi operators $c_{ia\sigma}$ with the
transformed fermions $\tilde{c}_{ia\sigma}$.

Expanding Eq. \ref{ut} yields
\be
H=\tilde{H} + [S,\tilde{H}] + \frac{1}{2}[S,[S,\tilde{H}]] + \cdots .
\label{newH}
\ee
$S$ is designed to satisfy Eq. \ref{constraint} order by order in perturbation theory
\be
S=S_1 + S_2 + \cdots,
\ee
which are all written in terms of dressed fermionic degrees of freedom.
The first-order $S_1$ solves
\be
[[S_1,\tilde{H}_I],\tilde{H}_I] = [-\tilde{H}_K,\tilde{H}_I].
\ee
A naive solution could be $[S_1,\tilde{H}_I] = -\tilde{H}_K$, but further simplifications can be made as explained below.
In principle one can always classify $\tilde{H}_K$ according to the change of the interaction energy 
for the corresponding hopping processes.
Hopping processes conserving the interaction energy, which is denoted as
$\tilde{H}^0_K$, commute with $\tilde{H}_I$ and thus produce no important effects in this perturbation expansion.
As a result, the suitable choice for $S_1$ should be
\be
[S_1,\tilde{H}_I] = -(\tilde{H}_K - \tilde{H}^0_K).
\label{s1}
\ee

To coarse-grain $\tilde{H}_K$, we use the following projection operators,
\bea
P_{i\alpha}^n(\beta,\gamma,\cdots) = \tilde{n}_{i\beta} \tilde{n}_{i\gamma}...\Pi_{\tau\neq\alpha\neq\beta\neq\gamma\cdots} (1-\tilde{n}_{i\tau}),
\label{pr}
\eea
where the Greek letters refer to $(a,\sigma)$, where $a=(1,2)$ refers to the orbital degrees of freedom, and $\alpha\neq\beta\neq\gamma\cdots$.
$P_{i\alpha}^n(\beta,\gamma,\cdots)$ projects out wavefunctions which
have $n$ 
electrons occupying states $(\beta,\gamma,\cdots)$ 
and no electrons in the remaining states on site $i$ except $\alpha$.
With the help of these projection operators, we can rewrite $\tilde{H}_K$ as
\bea
\tilde{H}_K &=& - \sum_{\langle ij\rangle a b\sigma} t_{ab}\sum_{n,m=0}^{2n_o-1}\sum_{(\beta,\gamma,...)}\sum_{(\beta',\gamma',\cdots)}\nn\\
&&P^n_{i(a,\sigma)}(\beta,\gamma,\cdots)\tilde{c}^\dagger_{ia\sigma}\tilde{c}_{jb\sigma}P^m_{j(b,\sigma)}(\beta',\gamma',\cdots). \nn\\
\label{prhk}
\eea
The final step is to classify the terms in Eq. \ref{prhk} according to
the change of interaction energy $E_I\neq 0$. This can be determined 
by the commutation relation, 
\be
[\tilde{H}_{K,E_I},\tilde{H}_I] = - E_I \tilde{H}_{K,E_I}.
\label{eicomm}
\ee
Substituting Eqs. \ref{eicomm} into \ref{s1}, we can express $S_1$ in the general form 
\be
S_1 = \sum_{E_I} \frac{\tilde{H}_{K,E_I}}{E_I}.
\ee

Although the above formalism is quite general, it is cubersome to write out all the terms explicitly for $n_o$-orbital systems.
In particular, if we are interested in the low energy effective theory, only parts of $S_1$ are necessary depending on the filling.
In the next section, we will present the results for first-order corrections to the intensity of the lower energy bands, $m^h(n)$, 
for filling $n=1-x$ as an example.

\subsection{Intensity of lower energy bands for $n=1-x$ in a two-orbital Mott system}
\label{dswt}
Terms that must be retained in $S_1$ for $n=1-x$ are those associated
with hopping processes that create or destroy doubly occupied sites, namely
\bea
\tilde{H}'_K = - \sum_{\langle ij\rangle a b\sigma} t_{ab} \sum_{\beta}P^1_{i(a,\sigma)}(\beta)\tilde{c}^\dagger_{ia\sigma}\tilde{c}_{jib\sigma}P^0_{j(b,\sigma)} + h.c.\nn\\
\eea
The evaluation of the commutator $[\tilde{H}'_K$,$\tilde{H}_I]$ is
complicated by Hund's coupling and the paired hopping.   The 
recipe for treating the case with a non-zero $J$ is as follows.
We consider the term
\be
-t_1 P^1_{i(1\uparrow)}\big[(1\downarrow)\big] \tilde{c}^\dagger_{i1\uparrow}\tilde{c}_{j1\uparrow} P^0_{j(1\uparrow)}
\label{1u1d}
\ee
which describes the hopping process of $\vert i 1\downarrow, j1\uparrow\rangle \to \vert i 1\downarrow 1\uparrow, j0\rangle$. 
Because $\vert i 1\downarrow 1\uparrow\rangle$ is not an eigenstate of $\tilde{H}_I$ (see Table \ref{table:one}), a commutator of the form in Eq. \ref{eicomm} 
can not be obtained.
Instead, we need to introduce
\bea
A^\pm_{ij}\equiv &&P^1_{i(1\uparrow)}\big[(1\downarrow)\big] \tilde{c}^\dagger_{i1\uparrow}\tilde{c}_{j1\uparrow} P^0_{j(1\uparrow)}\nn\\
&&\pm
\tilde{c}^\dagger_{i2\downarrow}\tilde{c}_{i1\downarrow}P^1_{i(2\uparrow)}\big[(1\downarrow)\big]\tilde{c}^\dagger_{i2\uparrow}\tilde{c}_{j1\uparrow}P_{j(1\uparrow)}^0.
\eea
Straightforward calculations give
\bea
&&[A^\pm_{ij},\tilde{H}_I]\vert \psi\rangle = \big(A_{ij}^\pm\tilde{H}_I- \tilde{H}_I A_{ij}^\pm\big)\vert \psi\rangle\nn\\
&=& \big(A_{ij}^\pm\tilde{H}_I- \tilde{H}_I A^\pm_{ij}\big)\vert i1\downarrow,j1\uparrow\rangle\otimes\vert ...\rangle\nn\\
&=& - \tilde{H}_I\big[\vert i1\downarrow 1\uparrow,j0\rangle \pm \vert i2\downarrow 2\uparrow,j0\rangle\big]\otimes\vert ...\rangle\nn\\
&=&-(U\pm J) \big[\vert i1\downarrow 1\uparrow,j0\rangle \pm \vert i2\downarrow 2\uparrow,j0\rangle\big]\otimes\vert ...\rangle\nn\\
&=&-(U\pm J) A^\pm_{ij}\vert \psi\rangle.
\label{apm}
\eea
In the above equations, we have used the fact that $A_{ij}$ only changes states related to sites $i$ and $j$ in $\vert \psi\rangle$
so that we just need to focus on the change of interaction energy on sites $i$ and $j$.
Combining Eqs. \ref{s1} and \ref{apm}, we find
\be
[\frac{-t_1}{2}\big(\frac{A^+_{ij}}{U+J}+\frac{A^-_{ij}}{U-J}\big),\tilde{H}_I]=
t_1 P^1_{i(1\uparrow)}\big[(1\downarrow)\big] \tilde{c}^\dagger_{i1\uparrow}\tilde{c}_{j1\uparrow} P^0_{j(1\uparrow)}.
\ee
Following the same procedure, we organize all the terms in the first-order correction associated with $\tilde{H}'_K$ as
\bea
S'_1 &=& \sum_{l=1}^4\frac{s_{1,E_l}-s_{1,-E_l}}{E_l},
\eea
where $E_1=U'-J$, $E_2=U'+J$, $E_3=U-J$, $E_4=U+J$, and $s_{1,-E_l} = s_{1,E_l}^\dagger$.
Detailed expressions for $s_{1,E_l}$ are given in Appendix A.

With these results in hand, we are now ready to evaluate the
intensity of the lower energy bands.  
The unitary transformation 
\be
c^\dagger_{ia\sigma} = \tilde{c}^\dagger_{ia\sigma} + \big[S_1,\tilde{c}^\dagger_{ia\sigma}\big] + \cdots
\label{fullc1}
\ee
relates the bare electrons to the transformed electron operators.
To derive the spectral properties, we need to separate
Eq. \ref{fullc1} into operators which are coarse-grained relative to
the energy scale.  
In particular, the zero energy component yields
\be
c^\dagger_{ia\sigma,0} = \tilde{c}^\dagger_{ia\sigma,0} - \big(\sum_{l=1}^4\frac{1}{E_l}\big[s_{1,-E_l},\tilde{c}^\dagger_{ia\sigma,E_l}\big]\big), \nn\\
\ee
where
\bea
\tilde{c}^\dagger_{ia\sigma,0} &=& \tilde{c}^\dagger_{ia\sigma}P^0_{ia\sigma},\nn\\
\tilde{c}^\dagger_{ia\sigma,U'-J} &=& \tilde{c}^\dagger_{ia\sigma}P^1_{ia\sigma}[(b\sigma)] + \tilde{c}^\dagger_{ia\sigma}P^1_{ia\sigma}[(b\bar{\sigma})]+
\tilde{c}^\dagger_{ia\bar{\sigma}}\tilde{c}^\dagger_{ib\sigma}\tilde{c}_{ib\bar{\sigma}},\nn\\
\tilde{c}^\dagger_{ia\sigma,U' + J} &=& \tilde{c}^\dagger_{ia\sigma}P^1_{ia\sigma}[(b\bar{\sigma})]- 
\tilde{c}^\dagger_{ia\bar{\sigma}}\tilde{c}^\dagger_{ib\sigma}\tilde{c}_{ib\bar{\sigma}},\nn\\
\tilde{c}^\dagger_{ia\sigma,U\mp J} &=& \tilde{c}^\dagger_{ia\sigma}P^1_{ia\sigma}[(a\bar{\sigma})]\mp
\tilde{c}^\dagger_{ib\sigma}\tilde{c}^\dagger_{ib\bar{\sigma}}\tilde{c}_{ia\bar{\sigma}},
\eea
 and $b\neq a$. 
Using the approximate expression for the intensity of the lower energy bands derived from the first moment of the spectral functions\cite{harris1967}, we have
\bea
m^h(1-x) &=&\frac{1}{{\cal N}}\sum_{ia\sigma}\langle\{c_{ia\sigma,0},c^\dagger_{ia\sigma,0}\}\rangle\nn\\
&=&\frac{1}{{\cal N}}\sum_{ia\sigma} 
\langle\{\tilde{c}_{ia\sigma,0},\tilde{c}^\dagger_{ia\sigma,0}\}\rangle\nn\\
&-& \langle\sum_{n=1}^4\big(\frac{1}{E_n}\big\{\big[\tilde{c}_{ia\sigma,E_n},s_{1,E_n}\big],\tilde{c}_{ia\sigma,0}^\dagger\big\}+h.c.\big)\rangle + \cdots.\nn\\
\label{mlhb}
\eea
While the first term in Eq. \ref{mlhb} yields $1+3x$ which is simply the spectral weight transfered in the atomic limit given by Eq. \ref{mlhb-1-x},
the second term in Eq. \ref{mlhb} is precisely the DSWT in question.
This expression can be simplified using two relationships.  First, 
$\tilde{c}_{ia\sigma,E}\vert G\rangle = \langle G\vert
\tilde{c}_{ia\sigma,E}^\dagger = 0$ since we are considering the case with filling $n=1-x\leq 1$.
Second,
$\tilde{c}_{ia\sigma,0}\tilde{c}_{ia\sigma,E}=\tilde{c}_{ia\sigma,0}\tilde{c}^\dagger_{ia\sigma,E}=0$. 
Applying these properties to Eq. \ref{mlhb} leads to
\be
\langle \big\{\big[\tilde{c}_{ia\sigma,E_n},s_{1,E_n}\big],\tilde{c}_{ia\sigma,0}^\dagger\big\}\rangle
=\langle\tilde{c}_{ia\sigma,E}s_{1,E}\tilde{c}^\dagger_{ia\sigma,0}\rangle,
\ee
and a straightforward calculation yields
\begin{widetext}
\bea
-\langle\tilde{c}_{ia\sigma,U'-J}s_{1,U'-J}\tilde{c}_{ia\sigma,0}^\dagger\rangle
&=&\sum_\delta t_b \langle \tilde{c}^\dagger_{ib\sigma}\tilde{c}_{i+\delta,b\sigma}\rangle
+t_{12} \langle \tilde{c}^\dagger_{ib\sigma}\tilde{c}_{i+\delta,a\sigma}\rangle 
+\sum_\delta
\frac{t_b}{2}\langle \tilde{c}^\dagger_{ib\bar{\sigma}}\tilde{c}_{i+\delta,b\bar{\sigma}} \rangle
+\frac{t_{12}}{2}\langle \tilde{c}^\dagger_{ib\bar{\sigma}}\tilde{c}_{i+\delta,a\bar{\sigma}}\rangle,\nn\\
-\langle\tilde{c}_{ia\sigma,U'+J}s_{1,U'+J}\tilde{c}_{ia\sigma,0}^\dagger\rangle
&=&\sum_\delta
\frac{t_b}{2}\langle \tilde{c}^\dagger_{ib\bar{\sigma}}\tilde{c}_{i+\delta,b\bar{\sigma}} \rangle
+\frac{t_{12}}{2}\langle \tilde{c}^\dagger_{ib\bar{\sigma}}\tilde{c}_{i+\delta,a\bar{\sigma}}\rangle,\nn\\
-\langle\tilde{c}_{ia\sigma,U-J}s_{1,U-J}\tilde{c}_{ia\sigma,0}^\dagger\rangle
&=&\sum_\delta
\frac{t_a}{2}\langle \tilde{c}^\dagger_{ia\bar{\sigma}}\tilde{c}_{i+\delta,a\bar{\sigma}}\rangle
+\frac{t_{12}}{2}\langle \tilde{c}^\dagger_{ia\bar{\sigma}}\tilde{c}_{i+\delta,b\bar{\sigma}}\rangle,\nn\\
-\langle\tilde{c}_{ia\sigma,U+J}s_{1,U+J}\tilde{c}_{ia\sigma,0}^\dagger\rangle
&=&\sum_\delta
\frac{t_a}{2}\langle \tilde{c}^\dagger_{ia\bar{\sigma}}\tilde{c}_{i+\delta,a\bar{\sigma}}\rangle
+\frac{t_{12}}{2}\langle \tilde{c}^\dagger_{ia\bar{\sigma}}\tilde{c}_{i+\delta,b\bar{\sigma}}\rangle,
\label{ciasigma}
\eea
\end{widetext}
where $\delta=\hat{x},\hat{y}$.
Substituting Eq. \ref{ciasigma} into Eq. \ref{mlhb}, we finally arrive at
\bea
m^h(1-x) &=& 1 + 3x + \alpha,\nn\\
\alpha&=& 2\big(\frac{1}{U'-J} + \frac{U'}{U'^2-J^2} + \frac{U}{U^2-J^2}\big) h_K,\nn\\
\eea
where $\alpha$ is a measure of the DSWT with
\bea
h_K &=& \frac{1}{{\cal N}}\sum_{<i,j>,\sigma}\big[t_1\langle \tilde{c}^\dagger_{i1\sigma}\tilde{c}_{j1\sigma}\rangle
+ t_2\langle \tilde{c}^\dagger_{i2\sigma}\tilde{c}_{j2\sigma}\rangle\nn\\ 
&+& t_{12}\big(\langle \tilde{c}^\dagger_{i1\sigma}\tilde{c}_{j2\sigma}\rangle+\langle \tilde{c}^\dagger_{i2\sigma}\tilde{c}_{j1\sigma}\rangle\big)\big],
\eea
the average kinetic energy of the empty site created in the lower band per site evaluated in the ground state. Note the presence of transformed operators in the expression for $h_K$.
This result represents a generalization of the single-band Hubbard
result\cite{eskes1994,harris1967} to the two-band case in the presence of Hund's coupling. 
It can be seen that DSWT is enahnced in the two-orbital system because the prefactor of $h_K$ in $\alpha$ shifts from $2/U$ in the single-band Hubbard model 
to $2\big(\frac{1}{U'-J} + \frac{U'}{U'^2-J^2} + \frac{U}{U^2-J^2}\big)$.
The physics behind this trend is as follows.
Since DSWT is induced at the expense of creating doubly occupied sites, more orbitals provide more ways to create such sites, which 
results in the enhancement of DSWT in a multi-orbital case.

\subsection{Effective low energy Hamiltonian for $n=1-x$}
While it is certainly complicated, we can also derive the effective
Hamiltonian in the transformed basis.  Eq. \ref{newH} leads to the effective Hamiltonian
\be
H=\tilde{H}_K + \tilde{H}_I + \big[S_1,\tilde{H}_K\big] + \cdots.
\label{effH}
\ee
Given the explicit expression for $S_1$ for $n=1-x$, one readily finds that
\be
H^{eff}_{n=1-x}=\tilde{H}'_K  + \sum_{b\neq a}\sum_{i,\delta,\delta',\sigma} \big(\frac{H_1+H_4}{U'-J} + \frac{H_2+H_5}{U'^2-J^2} + \frac{H_3+H_6}{U^2-J^2} + h.c.\big),
\ee
where
\begin{widetext}
\bea
H_1 &=& -t_a^2 P^0_{i+\delta',(a\sigma)}P^0_{i+\delta,(a\sigma)} 
\tilde{c}^\dagger_{i+\delta',a\sigma}\tilde{c}_{i+\delta,a\sigma}P^1_{i,(a\sigma)}[(b\sigma)]
-t_at_b P^0_{i+\delta',(b\sigma)}\tilde{c}^\dagger_{i+\delta',b\sigma}\tilde{c}_{ib\sigma}P^1_{i,(a\sigma)}[(b\sigma)]
\tilde{c}^\dagger_{ia\sigma}\tilde{c}_{i+\delta,a\sigma}P^0_{i+\delta,(a\sigma)}\nn\\
H_2 &=& -U'\big\{t_a^2  P^0_{i+\delta',(a\sigma)}P^0_{i+\delta,(a\sigma)}
\tilde{c}^\dagger_{i+\delta',a\sigma}\tilde{c}_{i+\delta,a\sigma}P^1_{i,(a\sigma)}[(b\bar{\sigma})]
+t_at_b P^0_{i+\delta',(b\bar{\sigma})}\tilde{c}^\dagger_{i+\delta',b\bar{\sigma}}\tilde{c}_{ib\bar{\sigma}}P^1_{i,(a\sigma)}[(b\bar{\sigma})]
\tilde{c}^\dagger_{ia\sigma}\tilde{c}_{i+\delta,a\sigma}P^0_{i+\delta,(a\sigma)}\big\}\nn\\
&-&Jt_a^2  P^0_{i+\delta',(a\sigma)}\tilde{c}^\dagger_{i+\delta',a\sigma}\tilde{c}_{i,a\sigma}\tilde{c}^\dagger_{i,b\sigma}\tilde{c}_{i,b\bar{\sigma}}
\tilde{c}^\dagger_{i,a\bar{\sigma}}\tilde{c}_{i+\delta,a\sigma}P^0_{i+\delta,(a\sigma)}\nn\\
&-&J  t_at_b P^0_{i+\delta',(b\bar{\sigma})}\tilde{c}^\dagger_{i+\delta',b\bar{\sigma}}\tilde{c}_{ib\bar{\sigma}}\tilde{c}^\dagger_{i+\delta,b\sigma}\tilde{c}_{i+\delta,b\bar{\sigma}}
P^1_{i+\delta,(a\bar{\sigma})}[(b\bar{\sigma})]\tilde{c}^\dagger_{i+\delta,a\bar{\sigma}}\tilde{c}_{ia\sigma}  P^0_{i,(a\sigma)}\nn\\
H_3 &=& -U\big\{t_a^2 P^0_{i+\delta',(a\sigma)}P^0_{i+\delta,(a\sigma)}
\tilde{c}^\dagger_{i+\delta',a\sigma}\tilde{c}_{i+\delta,a\sigma}P^1_{i,(a\sigma)}[(a\bar{\sigma})]
+t_a^2 P^0_{i+\delta',(a\bar{\sigma})}\tilde{c}^\dagger_{i+\delta',a\bar{\sigma}}\tilde{c}_{ia\bar{\sigma}}P^1_{i,(a\sigma)}[(b\bar{\sigma})]
\tilde{c}^\dagger_{ia\sigma}\tilde{c}_{i+\delta,a\sigma}P^0_{i+\delta,(a\sigma)}\big\}\nn\\
&-&J t_a^2 P^0_{i+\delta',(a\sigma)}\tilde{c}^\dagger_{i+\delta',a\sigma}\tilde{c}_{i,a\sigma}\tilde{c}^\dagger_{i,b\bar{\sigma}}\tilde{c}_{i,a\bar{\sigma}}
\tilde{c}^\dagger_{i,v\sigma}\tilde{c}_{i+\delta,a\sigma}P^0_{i+\delta,(a\sigma)}\nn\\
&-& J t_a^2 P^0_{i+\delta',(a\bar{\sigma})}\tilde{c}^\dagger_{i+\delta',a\bar{\sigma}}\tilde{c}_{ia\bar{\sigma}}\tilde{c}^\dagger_{i+\delta,b\bar{\sigma}}\tilde{c}_{i+\delta,a\bar{\sigma}}
P^1_{i+\delta,(b\sigma)}[(a\bar{\sigma})]\tilde{c}^\dagger_{i+\delta,b\sigma}\tilde{c}_{ia\sigma}  P^0_{i,(a\sigma)}
\nn\\
H_4 &=&
-t_{12}^2 P^1_{i+\delta,(b\sigma)}[(a\sigma)]P^0_{i+\delta',(a\sigma)}\tilde{c}^\dagger_{i+\delta',a\sigma}\tilde{c}_{i+\delta,a\sigma}P^0_{i+\delta,(a\sigma)}
-t_at_{12}P^0_{i+\delta',(a\sigma)}\tilde{c}^\dagger_{i+\delta',a\sigma}\tilde{c}_{ib\sigma}P^1_{i,(a\sigma)}[(b\sigma)]
\tilde{c}^\dagger_{ia\sigma}\tilde{c}_{i+\delta,a\sigma}P^0_{i+\delta,(a\sigma)}\nn\\
H_5 &=& -U't_{12}^2 
 P^1_{i+\delta,(b\sigma)}[(a\bar{\sigma})]P^0_{i+\delta',(a\sigma)}\tilde{c}^\dagger_{i+\delta',a\sigma}\tilde{c}_{i+\delta,a\sigma}P^0_{i+\delta,(a\sigma)}\nn\\
 &-&U't_at_{12}P^0_{i+\delta',(a\bar{\sigma})}\tilde{c}^\dagger_{i+\delta',a\bar{\sigma}}\tilde{c}_{ib\bar{\sigma}}P^1_{i,(a\sigma)}[(b\bar{\sigma})]
 \tilde{c}^\dagger_{ia\sigma}\tilde{c}_{i+\delta,a\sigma}P^0_{i+\delta,(a\sigma)}
\nn\\
H_6 &=& -Ut_{12}^2
 P^1_{i+\delta,(b\sigma)}[(b\bar{\sigma})]P^0_{i+\delta',(a\sigma)}\tilde{c}^\dagger_{i+\delta',a\sigma}\tilde{c}_{i+\delta,a\sigma}P^0_{i+\delta,(a\sigma)}\nn\\
 &-&U t_at_{12}P^0_{i+\delta',(b\bar{\sigma})}\tilde{c}^\dagger_{i+\delta',b\bar{\sigma}}\tilde{c}_{ia\bar{\sigma}}P^1_{i,(a\sigma)}[(a\bar{\sigma})]
  \tilde{c}^\dagger_{ia\sigma}\tilde{c}_{i+\delta,a\sigma}P^0_{i+\delta,(a\sigma)}.\nn\\
\label{t2u}
\eea
\end{widetext}
In deriving this expression, we used the properties $c_{ia\sigma}P^1_{i(a\sigma)}[(b\sigma')]\vert G\rangle = 0$ and $\tilde{H}_I\vert G\rangle = 0$ 
since the ground state can only have singly-occupied sites for $n=1-x$.
In Eq. \ref{t2u}, three-site hopping terms can be identified with $\delta\neq\delta'$.
Moreover, terms with $\delta=\delta'$ represent generalized superexchange interactions between nearest-neighbor sites occuring not only between 
spin but also orbital degrees of freedom.
Several terms proportional to $J$ are induced as a result of the internal change of states on the same sites allowed by the Hund's coupling and paired hopping.
As was the case with the single-band Hubbard model, this effective model contains three-site terms which can not be obtained by simply considering generalized superexchange processes between 
spin and orbital degrees of freedom.

\subsection{Particle-hole symmetric Quantum critical point}
Recently, it was proposed\cite{honma2008} that the thermopower offers an unbiased
scale for determining the doping level in the cuprates. The basis for
this observation is that the thermopower for a wide variety of
hole-doped cuprates all collapse onto a single curve and crosses zero\cite{honma2008}
at a universal doping level of $x=0.23$.  Since there is only one
fitting parameter in the thermpower and it is the doping, the
thermopower can be used to map out\cite{honma2008} the high-$T_c$ phase diagram. 
Physically, the thermo-electric power $S$ is defined as the voltage drop across a material generated by a temperature gradient.
Since itinerant carriers flow from higher to lower temperature
regardless of their charge, the vanishing of S occurs when particle-hole symmetry is present.

In Fermi liquid theory, particle-hole symmetry only holds at half-filling and any finite doping of holes should break this symmetry.
As a result, the explanation for a vanishing of S at optimal doping requires the incorporation of strong correlation effects.
One of us\cite{chakraborty2010b,phillips2010} has addresed this issue by considering the spectral weight transfer in the single-band Hubbard model in the strong coupling limit.
The particle-hole symmetric condition can be defined where the number of states below and above the chemical potential becomes equal.
In the atomic limit, this condition yields
\be
1-x_c=2 x_c,
\ee
which predicts the vanishing of $S$ at $x_c=1/3$. 
Away from the atomic limit, DSWT renders $x_c$ to a lower value around $x_c=0.23$ as observed in the experiments.  That particle-hole symmetry is dynamically generated near optimal in strong coupling models of the cuprates has also been confirmed by Vidhyadhiraja and co-workers\cite{vidhyadhiraja2009}.

This senario can be generalized to the current $n_o$-orbital Mott
system, allowing us to make the following predictions.
We present the particle-hole symmetric condition in the atomic limit to show how the critical doping with vanishing $S$ depends on the multi-orbital structure,
and it is expected that in the real materials this critical doping should be reduced to lower values due to DSWT.

In the case of $n=1-x$, the particle-hole symmetric condition is
unambiguously given by
\be\label{holds}
1-x_c = 2n_ox_c \Rightarrow x_c = \frac{1}{2n_o+1}
\ee
in the atomic limit which is strictly less than $1/5$ as would be the critical doping in the 2-orbital case.
Interestingly, Mukerjee\cite{mukerjee2005} has analyzed the thermopower in the atomic limit of the 3-orbital Hubbard model and found it to vanish at the doping level predicted by Eq. (\ref{holds}). This agreement further corroborates our treatment here that the particle-hole symmetric condition in the lower band shifts to lower values of doping as multi-orbital contributions are included.
For $n>1$, again the situation is complicaed by Hund's coupling and the paired hopping, and the particle-hole symmetric condition depends crucially on 
the magnitude of $J$.
If $J$ is large enough to produce new gaps in addition to those created by $U$, the particle-hole symmetric condition 
should be determined only by the bands in the vicinity of the chemical potential.
For example, for the case of $n=1+x$ shown in Fig. \ref{fig:one}(c), the particle-hole symmetric condition should be 
\be
2x_c=1.5(1-x_c)\Rightarrow x_c=\frac{3}{7}
\ee
if $J$ is large enough.
If $J$ is not large enough to generate new gaps, all the states within $\mu-E^-$ and $\mu+E^+$ defined in Eqs. \ref{mlhb-h} and \ref{mlhb-e} 
should contribute to the thermopower.
Consequently, in the latter case the particle-hole symmetric conditions can be written in general form
\be
n_i(1-x)= (2n_o-n_i+1)x \Rightarrow x_c^h(n_i) = \frac{n_i}{2 n_o + 1}
\label{xch}
\ee
for $n=n_i - x$, and
\be
(n_i+1)x = (2n_o - n_i)(1-x) \Rightarrow x_c^e(n_i) = \frac{2 n_o - n_i}{2 n_o + 1}
\label{xce}
\ee
for $n=n_i + x$.

These relationships have interesting consequences for superconductivity. 
Since the paricle-hole symmetric critical doping is intimately related
to the optimal doping of
superconductivity\cite{honma2008,chakraborty2010b,phillips2010}, we
can draw the following conclusion.  Optimal doping for
superconductivity in multi-orbital Mott
systems should be reduced relative to their single-orbital (that is, one
unpaired d electron) counterparts because the particle-hole symmetric
conditions derived above are all bounded from above by the
single-orbital value of $x=1/3$. 
According to Eqs. \ref{xch} and \ref{xce}, either a hole-doped multi-orbital Mott material with smaller integer fillings, or an electron-doped Mott multi-orbital material 
with larger integer fillings could serve as a good parent compound for new high-$T_c$ superconductors.

\section{Conclusions}
In this paper, spectral properties and related predictions have been studied in multi-orbital Mott systems for the first time. 
Taking a two-orbital system with full on-site interaction as a starting point, we have analyzed exact spectral functions in the atomic limit near integer 
fillings of $n_i=1,2,3$ and generalized the results to $n_o$ orbitals at arbitrary filling.  Most striking is the result that for $n=1-x$, the static part of the 
spectral weight of the unnoccupied part of the lower band grows as $4x$ as opposed to $2x$ in the single-orbital case.  This result can be easily be confirmed by 
doping a multi-orbital Mott system such as NiO and measuring the oxygen K-edge x-ray intensity as has been done previously for the cuprates\cite{chen1991}.
As the hopping terms are turned on, the redistribution of the spectral weight, namely dynamical spectral weight transfer (DSWT), occurs at all the energy 
scales in a manner similar to that in the single-band Hubbard model.

We have developed a general formalism to perform the perturbation theory in the strong coupling limit and applied this method to the two-orbital system.
The intensity of the lower energy bands has been explicitly calculated for filling $n=1-x$, and we have found that the leading-order correction due to DSWT is 
proportional to the average kinetic energy of the empty sites.
Relative to the single-band Hubbard model, we have shown that both static and dynamical spectral weight transfers are greatly enhanced as a consequence of 
multi-orbital structure.
The corresponding low energy effective Hamiltonian for $n=1-x$ in the two-orbital Mott system has been derived.
Moreover, we have shown that a critical doping, defined as the doping level exhibiting zero thermopower, is reduced to a smaller value as the number of orbitals increases. 
Motivated by the lesson from the cuprates that this critical doping is intimately related to the optimal doping of superconductivity,
we have proposed that either a hole-doped multi-orbital Mott material with smaller integer fillings, or an electron-doped Mott multi-orbital material
with larger integer fillings could serve as a good parent compound for new high-$T_c$ superconductors.

\section{Acknowledgements}
We would like to thank Seuming Hong and Weicheng Lv for help discussions. We acknowledge financial support from the NSF DMR-0940992 and the 
Center for Emergent Superconductivity, a DOE Energy Frontier Research Center, Grant No. DE-AC0298CH1088.

\section{Appendix A}
In this appendix, we write out $S'_1$ explicitly following the same procedure described in the Section \ref{dswt}.
To simplify the notation, we use a short-hand index of $1\to (1,\uparrow)$, $2\to (1,\downarrow)$, $3\to (2,\uparrow)$, $4\to (2,\downarrow)$.
This leads to the following result:
\bea
A_{ij}^\pm&\equiv& P^1_{i1}(2) \tilde{c}^\dagger_{i1}\tilde{c}_{j1} P^0_{j1}\pm \tilde{c}^\dagger_{i4}\tilde{c}_{i2}P^1_{i3}(2)\tilde{c}^\dagger_{i3}\tilde{c}_{j1}P_{j1}^0,\nn\\
&&\big[\frac{1}{2}\big(\frac{A^+_{ij}}{U+J}+\frac{A^-_{ij}}{U-J}\big),\tilde{H}_I\big] = - P^1_{i1}(2)\tilde{c}^\dagger_{i1}\tilde{c}_{j1} P^0_{j1},\nn
\eea
\bea
B_{ij}^\pm&\equiv& P^1_{i2}(1) \tilde{c}^\dagger_{i2}\tilde{c}_{j2} P^0_{j2}\pm \tilde{c}^\dagger_{i3}\tilde{c}_{i1}P^1_{i4}(1)\tilde{c}^\dagger_{i4}\tilde{c}_{j2}P_{j2}^0,\nn\\
&&\big[\frac{1}{2}\big(\frac{B_{ij}^+}{U+J}+\frac{B_{ij}^-}{U-J}\big),\tilde{H}_I\big] = - P^1_{i2}(1) \tilde{c}^\dagger_{i2}\tilde{c}_{j2} P^0_{j2},\nn
\eea
\bea
C_{ij}^\pm&\equiv& P^1_{i3}(4) \tilde{c}^\dagger_{i3}\tilde{c}_{j3} P^0_{j3}\pm \tilde{c}^\dagger_{i2}\tilde{c}_{i4}P^1_{i1}(4)\tilde{c}^\dagger_{i1}\tilde{c}_{j3}P_{j3}^0,\nn\\
&&\big[\frac{1}{2}\big(\frac{C_{ij}^+}{U+J}+\frac{C_{ij}^-}{U-J}\big),\tilde{H}_I\big] = - P^1_{i3}(4) \tilde{c}^\dagger_{i3}\tilde{c}_{j3} P^0_{j3},\nn
\eea
\bea
D_{ij}^\pm&\equiv& P^1_{i4}(3) \tilde{c}^\dagger_{i4}\tilde{c}_{j4} P^0_{j4}\pm \tilde{c}^\dagger_{i1}\tilde{c}_{i3}P^1_{i2}(3)\tilde{c}^\dagger_{i2}\tilde{c}_{j4}P_{j4}^0,\nn\\
&&\big[\frac{1}{2}\big(\frac{D_{ij}^+}{U+J}+\frac{D_{ij}^-}{U-J}\big),\tilde{H}_I\big] = - P^1_{i4}(3) \tilde{c}^\dagger_{i4}\tilde{c}_{j4} P^0_{j4},\nn
\eea
\bea
E_{ij}^\pm&\equiv& P^1_{i3}(4) \tilde{c}^\dagger_{i3}\tilde{c}_{j1} P^0_{j1}\pm \tilde{c}^\dagger_{i2}\tilde{c}_{i4}P^1_{i1}(4)\tilde{c}^\dagger_{i1}\tilde{c}_{j1}P_{j1}^0,\nn\\
&&\big[\frac{1}{2}\big(\frac{E_{ij}^+}{U+J}+\frac{E_{ij}^-}{U-J}\big),\tilde{H}_I\big] = - P^1_{i3}(4) \tilde{c}^\dagger_{i3}\tilde{c}_{j1} P^0_{j1},\nn
\eea
\bea
F_{ij}^\pm&\equiv& P^1_{i4}(3) \tilde{c}^\dagger_{i4}\tilde{c}_{j2} P^0_{j2}\pm \tilde{c}^\dagger_{i1}\tilde{c}_{i3}P^1_{i2}(3)\tilde{c}^\dagger_{i2}\tilde{c}_{j2}P_{j2}^0,\nn\\
&&\big[\frac{1}{2}\big(\frac{F_{ij}^+}{U+J}+\frac{F_{ij}^-}{U-J}\big),\tilde{H}_I\big] = - P^1_{i4}(3) \tilde{c}^\dagger_{i4}\tilde{c}_{j2} P^0_{j2},\nn
\eea
\bea
G_{ij}^\pm&\equiv& P^1_{i1}(2) \tilde{c}^\dagger_{i1}\tilde{c}_{j3} P^0_{j3}\pm \tilde{c}^\dagger_{i4}\tilde{c}_{i2}P^1_{i3}(2)\tilde{c}^\dagger_{i3}\tilde{c}_{j3}P_{j3}^0,\nn\\
&&\big[\frac{1}{2}\big(\frac{G_{ij}^+}{U+J}+\frac{G_{ij}^-}{U-J}\big),\tilde{H}_I\big] = - P^1_{i1}(2) \tilde{c}^\dagger_{i1}\tilde{c}_{j3} P^0_{j3},\nn
\eea
\bea
H_{ij}^\pm&\equiv& P^1_{i2}(1) \tilde{c}^\dagger_{i2}\tilde{c}_{j4} P^0_{j4}\pm \tilde{c}^\dagger_{i3}\tilde{c}_{i1}P^1_{i4}(1)\tilde{c}^\dagger_{i4}\tilde{c}_{j4}P_{j4}^0,\nn\\
&&\big[\frac{1}{2}\big(\frac{H_{ij}^+}{U+J}+\frac{H_{ij}^-}{U-J}\big),\tilde{H}_I\big] = - P^1_{i2}(1) \tilde{c}^\dagger_{i2}\tilde{c}_{j4} P^0_{j4}.\nn
\eea

We can also apply the same technique to hopping terms leading to final states with $\vert i1\uparrow 2\downarrow, j0\rangle$ 
or $\vert i1\downarrow 2\uparrow,j0\rangle$, but this time 
there exists two different ways. 
For example, we can define,
\be
K_{ij}^\pm\equiv P^1_{i1}(4) \tilde{c}^\dagger_{i1}\tilde{c}_{j1} P^0_{j1}\pm \tilde{c}^\dagger_{i3}\tilde{c}_{i4}P^1_{i2}(4)\tilde{c}^\dagger_{i2}\tilde{c}_{j1}P_{j1}^0,
\label{k1}
\ee
or
\be
K_{ij}^\pm\equiv P^1_{i1}(4) \tilde{c}^\dagger_{i1}\tilde{c}_{j1} P^0_{j1}\mp \tilde{c}^\dagger_{i2}\tilde{c}_{i4}P^1_{i3}(4)\tilde{c}^\dagger_{i3}\tilde{c}_{j1}P_{j1}^0,
\label{k2}
\ee
Both choices can lead to,
\be
\big[\frac{1}{2}\big(\frac{K_{ij}^+}{U'-J}+\frac{K_{ij}^-}{U'+J}\big),\tilde{H}_I\big] = -P^1_{i1}(4) \tilde{c}^\dagger_{i1}\tilde{c}_{j1} P^0_{j1},
\ee
This is because for the transformation from $\vert 1\uparrow 2\downarrow\rangle$ to $\vert 1\downarrow 2\uparrow\rangle$, we can either flip the spin 
(Eq. \ref{k1}) or exchange the orbital (Eq. \ref{k2}). 
Since both choices are legitimate and related up to another unitary transformation, we can choose the spin-flip scenario for the hoppings with $t_1$ and $t_2$ and 
the orbital exchange scenario for the hoppings with $t_{12}$ in this category without loss of generality.
We find immediately that
\bea
K_{ij}^\pm&\equiv& P^1_{i1}(4) \tilde{c}^\dagger_{i1}\tilde{c}_{j1} P^0_{j1}\pm \tilde{c}^\dagger_{i3}\tilde{c}_{i4}P^1_{i2}(4)\tilde{c}^\dagger_{i2}\tilde{c}_{j1}P_{j1}^0,\nn\\
&&\big[\frac{1}{2}\big(\frac{K_{ij}^+}{U'-J}+\frac{K_{ij}^-}{U'+J}\big),\tilde{H}_I\big] = -P^1_{i1}(4) \tilde{c}^\dagger_{i1}\tilde{c}_{j1} P^0_{j1},\nn
\eea
\bea
L_{ij}^\pm&\equiv& P^1_{i2}(3) \tilde{c}^\dagger_{i2}\tilde{c}_{j2} P^0_{j2}\pm \tilde{c}^\dagger_{i4}\tilde{c}_{i3}P^1_{i1}(3)\tilde{c}^\dagger_{i1}\tilde{c}_{j2}P_{j2}^0,\nn\\
&&\big[\frac{1}{2}\big(\frac{L_{ij}^+}{U'-J}+\frac{L_{ij}^-}{U'+J}\big),\tilde{H}_I\big] = -P^1_{i2}(3) \tilde{c}^\dagger_{i2}\tilde{c}_{j2} P^0_{j2},\nn
\eea
\bea
M_{ij}^\pm&\equiv& P^1_{i3}(2) \tilde{c}^\dagger_{i3}\tilde{c}_{j3} P^0_{j3}\pm \tilde{c}^\dagger_{i1}\tilde{c}_{i2}P^1_{i4}(2)\tilde{c}^\dagger_{i4}\tilde{c}_{j3}P_{j3}^0,\nn\\
&&\big[\frac{1}{2}\big(\frac{M_{ij}^+}{U'-J}+\frac{M_{ij}^-}{U'+J}\big),\tilde{H}_I\big] = -P^1_{i3}(2) \tilde{c}^\dagger_{i3}\tilde{c}_{j3} P^0_{j3},\nn
\eea
\bea
N_{ij}^\pm&\equiv& P^1_{i4}(1) \tilde{c}^\dagger_{i4}\tilde{c}_{j4} P^0_{j4}\pm \tilde{c}^\dagger_{i2}\tilde{c}_{i1}P^1_{i3}(1)\tilde{c}^\dagger_{i3}\tilde{c}_{j4}P_{j4}^0,\nn\\
&&\big[\frac{1}{2}\big(\frac{N_{ij}^+}{U'-J}+\frac{N_{ij}^-}{U'+J}\big),\tilde{H}_I\big] = -P^1_{i4}(1) \tilde{c}^\dagger_{i4}\tilde{c}_{j4} P^0_{j4},\nn
\eea
\bea
O_{ij}^\pm&\equiv& P^1_{i3}(2) \tilde{c}^\dagger_{i3}\tilde{c}_{j1} P^0_{j1}\mp \tilde{c}^\dagger_{i4}\tilde{c}_{i2}P^1_{i1}(2)\tilde{c}^\dagger_{i1}\tilde{c}_{j1}P_{j1}^0,\nn\\
&&\big[\frac{1}{2}\big(\frac{O_{ij}^+}{U'-J}+\frac{O_{ij}^-}{U'+J}\big),\tilde{H}_I\big] = -P^1_{i3}(2) \tilde{c}^\dagger_{i3}\tilde{c}_{j1} P^0_{j1},\nn
\eea
\bea
Q_{ij}^\pm&\equiv& P^1_{i4}(1) \tilde{c}^\dagger_{i4}\tilde{c}_{j2} P^0_{j2}\mp \tilde{c}^\dagger_{i3}\tilde{c}_{i1}P^1_{i2}(1)\tilde{c}^\dagger_{i2}\tilde{c}_{j2}P_{j2}^0,\nn\\
&&\big[\frac{1}{2}\big(\frac{Q_{ij}^+}{U'-J}+\frac{Q_{ij}^-}{U'+J}\big),\tilde{H}_I\big] = -P^1_{i4}(1) \tilde{c}^\dagger_{i4}\tilde{c}_{j2} P^0_{j2},\nn
\eea
\bea
R_{ij}^\pm&\equiv& P^1_{i1}(4) \tilde{c}^\dagger_{i1}\tilde{c}_{j3} P^0_{j3}\mp \tilde{c}^\dagger_{i2}\tilde{c}_{i4}P^1_{i3}(4)\tilde{c}^\dagger_{i3}\tilde{c}_{j3}P_{j3}^0,\nn\\
&&\big[\frac{1}{2}\big(\frac{R_{ij}^+}{U'-J}+\frac{R_{ij}^-}{U'+J}\big),\tilde{H}_I\big] = -P^1_{i1}(4) \tilde{c}^\dagger_{i1}\tilde{c}_{j3} P^0_{j3},\nn
\eea
\bea
W_{ij}^\pm&\equiv& P^1_{i2}(3) \tilde{c}^\dagger_{i2}\tilde{c}_{j4} P^0_{j4}\mp \tilde{c}^\dagger_{i1}\tilde{c}_{i3}P^1_{i4}(3)\tilde{c}^\dagger_{i4}\tilde{c}_{j4}P_{j4}^0,\nn\\
&&\big[\frac{1}{2}\big(\frac{W_{ij}^+}{U'-J}+\frac{W_{ij}^-}{U'+J}\big),\tilde{H}_I\big] = -P^1_{i2}(3) \tilde{c}^\dagger_{i2}\tilde{c}_{j4} P^0_{j4}.\nn
\eea
For hopping terms leading to final states with $\vert i1\sigma 2\sigma, j0\rangle$, no complication occurs because these final states are still eigenstates of $\tilde{H}_I$.
It is then straightforward to obtain
\bea
\tilde{T}_{U'-J} = \sum_{<i,j>} &-&t_1 \big[P^1_{i1}(3) \tilde{c}^\dagger_{i1} \tilde{c}_{j1}P^0_{j1} + P^1_{i2}(4) \tilde{c}^\dagger_{i2} \tilde{c}_{j2}P^0_{j2}\big] \nn\\
&-&t_2 \big[P^1_{i3}(1) \tilde{c}^\dagger_{i3} \tilde{c}_{j3}P^0_{j3} + P^1_{i4}(2) \tilde{c}^\dagger_{i4} \tilde{c}_{j4}P^0_{j4}\big] \nn\\
&-&t_{12} \big[P^1_{i3}(1) \tilde{c}^\dagger_{i3} \tilde{c}_{j1}P^0_{j1} + P^1_{i4}(2) \tilde{c}^\dagger_{i4} \tilde{c}_{j2}P^0_{j2}\nn\\
&+& P^1_{i1}(3) \tilde{c}^\dagger_{i1} \tilde{c}_{j3}P^0_{j3} + P^1_{i2}(4) \tilde{c}^\dagger_{i2} \tilde{c}_{j4}P^0_{j4}\big].\nn\\
\eea

Collecting all the above terms, for $n_f\leq 1$, we find that  the relevant terms in $S_1$ is
\bea
S'_1 &\equiv& \sum_{l=1}^4\frac{s_{1,E_l}-s_{1,-E_l}}{E_l},\nn\\
s_{1,U'-J} &=& \tilde{T}_{U'-J} + \sum_{<i,j>}-t_1(K_{ij}^+ +L_{ij}^+)-t_2(M_{ij}^+ +N_{ij}^+)\nn\\
&-&t_{12}(O_{ij}^++Q_{ij}^++R_{ij}^++W_{ij}^+)\nn\\
s_{1,U'+J} &=& \sum_{<i,j>}-t_1(K_{ij}^- +L_{ij}^-)-t_2(M_{ij}^- +N_{ij}^-)\nn\\
&-&t_{12}(O_{ij}^- +Q_{ij}^- +R_{ij}^- +W_{ij}^-)\nn\\
s_{1,U-J} &=& \sum_{<i,j>}-t_1(A_{ij}^- +B_{ij}^-)-t_2(C_{ij}^- +D_{ij}^-)\nn\\
&-&t_{12}(E_{ij}^-+F_{ij}^-+G_{ij}^-+H_{ij}^-)\nn\\
s_{1,U+J} &=& \sum_{<i,j>}-t_1(A_{ij}^++B_{ij}^+)-t_2(C_{ij}^++D_{ij}^+)\nn\\
&-&t_{12}(E_{ij}^++F_{ij}^++G_{ij}^++H_{ij}^+),\nn\\
\eea
where $E_1=U'-J$, $E_2=U'+J$, $E_3=U-J$, $E_4=U+J$, and $s_{1,-E_l} = s_{1,E_l}^\dagger$.

\end{document}